\documentclass{epl}

\title{Quenched noise and over-active sites in sandpile dynamics}
\shorttitle{Quenched noise... in sandpile dynamics}
\author{Mikko J. Alava\inst{1}\thanks{E-mail: \email{mja@fyslab.hut.fi}}
\and 	Kent B{\ae}kgaard Lauritsen\inst{2}\thanks{E:mail: \email{
baekgard@nbi.dk}}
}
\shortauthor{Mikko Alava \etal}
\institute{
  \inst{1} Helsinki University of Technology, Laboratory of Physics,
             02105 HUT, Finland\\
  \inst{2} Atmosphere Ionosphere Research Division,
           Danish Meteorological Institute, 2100 Copenhagen, Denmark
}

\begin{document}

\maketitle

\begin{abstract}
The dynamics of sandpile models are mapped to discrete interface equations.
We study in detail the Bak-Tang-Wiesenfeld model,
a stochastic model with random thresholds, and the
Manna model. These sandpile models are, respectively, discretizations of the
Edwards-Wilkinson equation with columnar, point-like
and correlated quenched noise, with the constraint that the interface velocity 
is either zero or one. The constraint, embedded in the
sandpile rules, gives rise to another noise component.
Studies of this term for the Bak-Tang-Wiesenfeld model reveal
long-range on-site correlations and that with open boundary conditions 
there is no spatial translational invariance. 
\end{abstract}

Simple cellular automata have been under intensive
study during the last decade. The aim has been
to explain many of the power-laws that can be often seen 
in systems in nature. A particularly famous example 
is the paradigmatic Bak-Tang-Wiesenfeld (BTW) sandpile model \cite{btw}, 
which exhibits both spatial and temporal criticality using
the language of statistical mechanics. Many 
variants of sandpile models have been developed, while
the BTW model still occupies a central position due to its
deceptive simplicity and complicated behavior \cite{dhar:1998}.

The central ingredients of a sandpile model 
are rules for redistributing grains and for
driving the system by grain addition if it is stable.
This gives rise to two time scales, slow and fast, and
usually to diffusive grain dynamics.
Grains are removed from unstable sites and distributed
to neighbors. Those lost through
the open boundaries are compensated by adding grains randomly, 
but on a slow time scale compared to the time that individual
avalanches or periods of activity take. In the asymptotic
self-organized critical (SOC) steady state power-laws
emerge in the avalanche statistics.
The scenario is reminiscent of another problem in
non-equilibrium statistical mechanics: driven interfaces
which couple a random environment to an elastic object.
At a critical value of the drive force the velocity
of the interface becomes non-zero, and exactly at this
depinning transition critical behavior ensues. The time and length
scales are renormalized, with associated critical exponents
\cite{narayan-fisher:1993,nattermann-etal:1992,leschhorn:1993}.

In this Letter we demonstrate a connection between interfaces
and many sandpile models. We construct a mapping of the
BTW model, and two stochastic models (``rice-pile'' models
\cite{paczuski-boettcher:1996,amaral-lauritsen}, the Manna-model
\cite{manna:1992}) to discrete models of driven interfaces \cite{earlier}. 
The interface field is constructed such that it counts topplings
as function of time. By employing the mapping, one obtains a description
of the stochastic movement of sandpile grains 
as an interface propagating through a
{\em quenched\/} random medium.
Such a description of sandpile dynamics 
allows us to address several issues in
SOC. First, the universality classes of different models are reflected
in the {\it noise terms\/} of the interface equation.
The quenched noise arises from the shot noise of grain deposition, and
this is the important feature in the BTW model. The noise can 
also result
from the fluctuations in the movement of grains that is
caused by randomness in the rules. Finally, 
disorder is created by the fluctuations in the density of 
active (unstable) sites. 
A grain may sometimes end on a site that would become 
anyways unstable. Such {\it over-active\/} sites correspond to interface
locations with a local velocity constrained to unity. 
This phenomenon is most distinctive in the BTW model, as
compared to the stochastic threshold/rice-pile and Manna models.
The rice-pile model is exactly a discrete version of 
the (random field) linear interface
model class and may be discussed with well-known
scaling and renormalization arguments 
\cite{narayan-fisher:1993,nattermann-etal:1992}. 
The mapping establishes 
the upper critical dimension to $d_u = 4$ for all the models.  
As a novel effect, the noise coming from the velocity constraint
demonstrates a lack of translational invariance
in open systems.

The sandpile models are as follows: each site $x$ of a
hyper-cubic lattice  of size $L^d$ has $z(x,t)$ grains.
When $z(x,t)$ exceeds a critical threshold
$z_c(x)$, the site is active and topples.
Grains are removed from $x$ and given 
to the nearest neighbors (nn). This in turn may cause
some nn's to topple and so on. If there are no active sites in
the system, one grain is added to a randomly chosen site, 
$z(x,t) \to z(x,t) + 1$.
The time and number of topplings till the system again contains no
active sites define an avalanche and its internal lifetime.
The system is usually open such that grains which topple out of the
system are lost (in $d=1$: $z_0 \equiv z_{L+1} \equiv 0$).
The specific toppling rule distinguishes between the models.
The BTW model has $z_c$ equal to a constant, $z_c=2d-1$
and the toppling rule:
$
        z(x,t+1) = z(x,t) - 2d, ~
        z(y,t+1) = z(y,t) + 1, 
$
where $y$ denotes all the $2d$ nn's of the site $x$.
The rice pile model has the same rule 
but with $z_c (x)$ randomly chosen after each toppling from
a probability distribution. In the Manna model, two
grains are given to two randomly chosen neighbors with $z_c =1$.
We consider here two kinds of ensembles. The SOC one is defined
above via the drive and the open boundaries. Another possibility
is to use periodic boundary conditions, and prepare the system
at a certain average 'energy' $\langle z(x,0)\rangle$ in which
case the critical state is reached at a certain energy
$\langle z \rangle_c$ \cite{dickman-etal:1998}.

The mapping of the dynamics of sandpile models \cite{earlier}. 
begins with the definition of an interface or memory field $H(x,t)$.
It counts topplings at site $x$ up to time $t$. 
The dynamics of $H$ defined in this way reads
\begin{eqnarray}
        H(x,t+1)&  = &\left\{\begin{array}{ll}
                        H(x,t) + 1,  &  f(x,t) >0  \\
                        H(x,t),      &  f(x,t) \le 0
                   \end{array} \right.
%                           \label{eq:H(x,t+1)}
 ~~~ \rightarrow ~~~
	\frac{\Delta H}{\Delta t}=\theta \left(f(x,t) \right) , 
                           \label{eq:H-def}
\end{eqnarray}
where we have rewritten the equation of motion for $H$ in the form of
a discrete interface equation. The important point
is the construction of the local ``force'' $f(x,t) = z(x,t)-z_{c}(x)$
at site $x$. 
We can express $f(x,t)$ in terms of the grains added to site
$x$, $n_x^{in}$, and removed from $x$, $n_x^{out}$, as
$f(x,t) = n_x^{in} - n_x^{out} - z_{c}(x)$, since 
$z(x,t) = n_x^{in} - n_x^{out}$. 

For the BTW and rice-pile models
$n_{x}^{in}$ and $n_{x}^{out}$ can be derived from
the local height field $H(x,t)$ and an external force term $F(x,t)$.
Here, $F$ counts the number of grains added to site $x$ 
by the external drive up to time $t$. Thus for SOC
drive $F(x,t)$ increases on the slow time scale and does not
change during avalanches, i.e.\ it acts
as a (columnar) quenched noise.
The construction of $f$ using $F$ works for any drive 
(e.g.\ continuous, uniform).
Observing that $n_x^{out}$ is simply $n_x^{out}=2d H(x,t)$
and $n_x^{in}=\sum_{x_{nn}} H(x_{nn},t)+F(x,t)$, where 
$x_{nn}$ denotes the
$2d$ nn's of $x$, one arrives at
$f(x,t) = \nabla^2 H + F(x,t) - z_c(x,H)$,
where $\nabla^2 H$ is the discrete Laplacian
\cite{unpublished,prep}.
The threshold $z_c(x,H)$ depends on $H$ for the rice-pile model
and reveals that the random thresholds correspond to
a quenched force field which is random in $x$ and $H$,
and acts on the interface. For the BTW model $z_c$ is constant. 

For the Manna model we have to incorporate
the randomness in the distribution rule. We
use a projection technique by writing $n_x^{in}$
as the average incoming flux $\bar{n}_x^{in}$ 
plus a fluctuating part $\delta n_x^{in}$ and arrive at
\begin{equation}
	f(x,t)  =  1/d \,\, \nabla^2 H + F(x,t) - z_c(x,H) + \tau(x,H),
							\label{force}
\end{equation}
with the resulting noise term
\begin{equation}
	\tau(x,H) = \delta n_x^{in}  \equiv  n_x^{in} - \bar{n}_x^{in}
	= n_x^{in}- 1/d \sum_{x_{nn}} H(x_{nn},t).
\end{equation}
Each nn-toppling contributes $1/d$ to $\bar{n}_x^{in}$
as the sum over the nn's is the expected number of grains from
neighboring sites due to their topplings. When site $x$ topples,
one uses the above definition of $\tau(x,H)$ in terms of the
fluxes ($n^{in}$, $\bar{n}^{in}$) to evaluate it at site $x$ and
height $H$. The projection trick, used to construct
the quenched noise $\tau$, means that the randomness in the toppling rule
is matched to an equivalent noise field $\tau$ so that
the interface equation for $H$ reproduces exactly the behavior
of the Manna sandpile. It can also be used in the case of
other models, where the effect of a toppling is random.
Note that $\tau(x,H)$ is a conserving noise since
the random toppling rule for the Manna model
conserves the number of grains (except at open boundaries).

\begin{figure}
\onefigure[width=7cm]{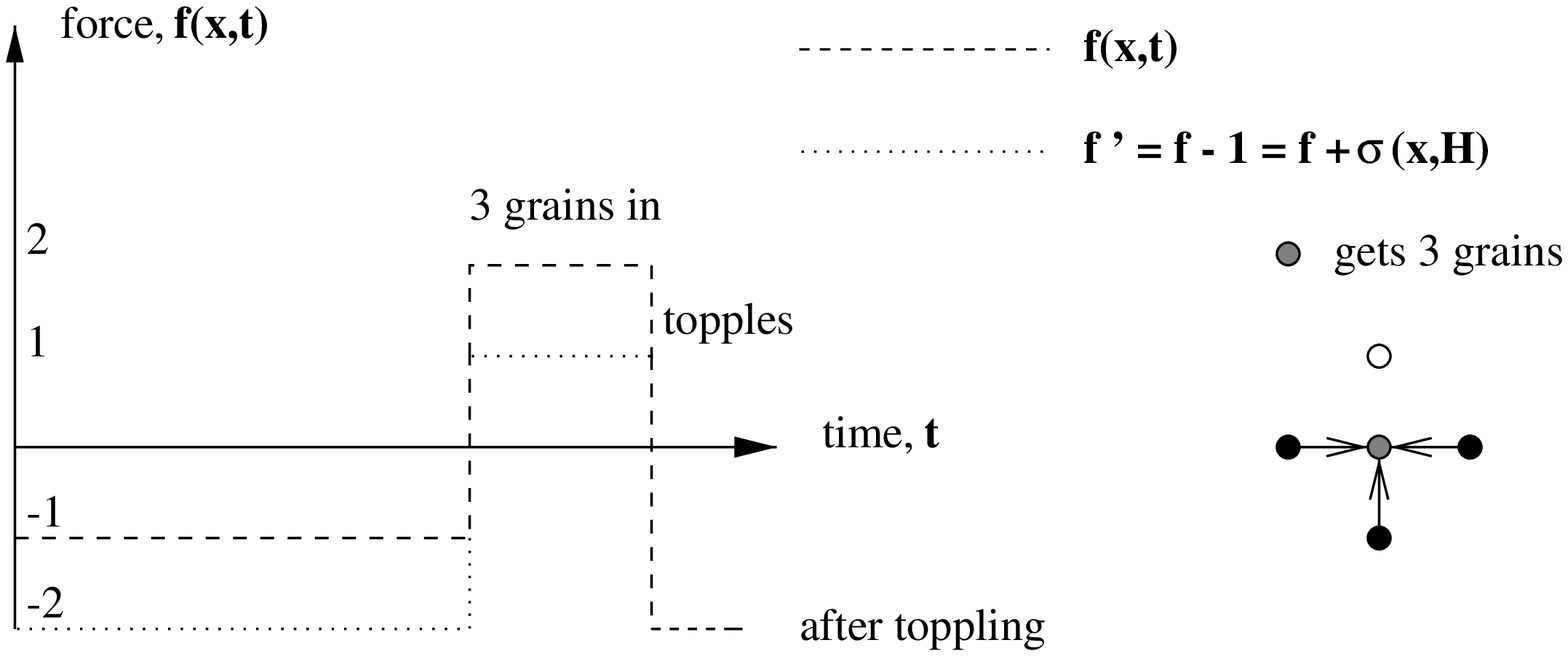}
\caption{Rescaling of the force $f$ and an example
of how the $\sigma$-noise ensues (three grains added simultaneously).}
\label{fig:explain}
\end{figure}

%Fig.1
The step-function, $\theta(f)$, in Equation~(\ref{eq:H-def}) 
forces it so that the interface does not move backwards and such that 
the velocity $v \equiv \Delta H / \Delta t$ is either 0 or 1
\cite{leschhorn:1993}. This
means that sandpiles are equivalent to cellular automaton models
of interface depinning, with this velocity constraint. We next map this
constraint into an effective noise term,
denoted $\sigma$, in the interface equation,
that can be used to discuss the possible differences of 'real'
depinning models and those that arise from sandpile models via
the mapping. 
Consider the toppling example in Fig.~\ref{fig:explain}.
On the avalanche timescale $f<0$ at site $x$.
As function of time, $f$ increases until at time $t-1$ several neighbors
topple resulting in $f>0$, so that site $x$ will topple at time $t$
and $\Delta H / \Delta t = 1$. 

The sandpile rules result in an {\it effective\/}
force $f' \equiv 1$ that acts on the interface $H$ at
the time of toppling:
$\Delta H/\Delta t \equiv f' \, \theta(f) = f' \theta(f')$.
The relation between $f$ and $f'$ for each toppling at $x$
(constant $H$) reads thus
\begin{equation}
	f'(x,t) = f(x,t) + \sigma(x,H) 
        ~~~ \rightarrow ~~~
       \sigma(x,H) = 1 + z_c(x,H) - z(x,t^*)
					\label{f-renorm}
\end{equation}
where $t^*$ is the time at which site $x$ topples such
that $\sigma(x,H)$ by this construction is a {\em quenched\/} random variable.
It is computed from the difference between $f$ and $f'$
when $x$ topples. Notice that
$f'$ and $f$ are by definition time-dependent variables, since
they change as grains are moved, or the Laplacian changes.
They also contain a quenched force component, as is seen from 
Eq.~(\ref{force}).
The easiest way to study the $\sigma$-noise is to
determine it in a simulation using Equation (\ref{f-renorm}).
This construction of $\sigma$ is similar to that of the 
$\tau$-noise term for the Manna-model and other models for
which the projection trick can be used.
The trick maps the difference of the expected value of
the local force at toppling $f(x,t^* ) = z_x - z_c$ and the true one to
$\tau$. Here the difference of $f'(x,t^* )\equiv 1$ 
and $f$ maps to $\sigma$, a quenched variable. Such differences arise
in the Manna model due to randomness in grain movement, 
and in the case of the $\sigma$-noise from the effect of the step-function.

The point in the noise variables $\sigma(x,H)$ and
$\tau(x,H)$ is that given these disorder fields, 
the interface equation exactly reproduces the history
or dynamics of a sandpile ``run''. Thus we can study
the interface model as such, and try to infer the
properties of the original sandpile from its 
behavior. Also, the $\sigma$-noise
allows one to study the effects of the peculiar discretization
($dH/dt = \theta(f)$) explicitly, since
$\Delta H/\Delta t = f' \theta(f' )$ can be interpreted as the 
discretization of the continuum equation $\partial H / \partial t = f'$ 
\cite{leschhorn:1993}. 
Combining Eqs.~(\ref{eq:H-def}) and (\ref{f-renorm})
we write the discretized interface equation as
\begin{equation}
	\frac{\Delta H}{\Delta t}  = 
	        \nu \nabla^2 H + \eta(x,H) + F(x,t) +\sigma(x,H) .
%	\nonumber\\
%	&& ~~~ \times \theta\left(\nu \nabla^2 H + \eta(x,H) + F(x,t) 
%                                      +\sigma(x,H)\right) ,
					\label{eq:btw-equ}
\end{equation}
Here the diffusion constant $\nu$ is unity for the BTW and ricepile
models and $1/d$ for the Manna model,
and the quenched noise $\eta(x,H) = - z_c(x,H) + \tau(x,H)$.
Equation~(\ref{eq:btw-equ}) is the central difference
discretization of a continuum diffusion equation
with quenched noise, called the linear interface
model (LIM) or the quenched Edwards-Wilkinson equation 
\cite{narayan-fisher:1993,nattermann-etal:1992,leschhorn:1993}.
Equation~(\ref{eq:btw-equ}) contains two main
ingredients. First, the Laplacian character of these sandpile
models which is such that the differences in accumulated
topplings map exactly to an elastic force. In the interface
language once the force increases sufficiently to overcome the pinning force,
$\nabla^2 H + \eta(x,H) + F(x,t)$,
 the interface moves by one step.
For open boundary sandpiles, the right interface boundary
condition is $H = 0$ which is to be imposed at ``extra sites''
($x=0$, $x=L+1$ for a system of size $L$ in 1d). In the SOC
steady-state the Laplacian increases, because of the roughly parabolic
shape for the the toppling profile $H(x)$ (see \cite{unpublished,barrat}).
This is compensated by the ever-increasing $\langle F(t) \rangle$,
or the addition of grains by the SOC drive.

Second, the randomness in the sandpile rules
map into noise variables [$F$, $\eta$],
as do the details of the dynamics [$\sigma$].
Equation~(\ref{eq:btw-equ}) allows thus to make conclusions
about the universality classes of models based on the noise
terms and their relevance. The $F$-term in Eq.~(\ref{eq:btw-equ}) 
is {\it columnar\/}:
it integrates the deposited grains and is constant during avalanches. 
The $\eta(x,H)$ term in the rice-pile and Manna models
explicitly depends on $H$. For the rice-pile 
$\eta(x,H) \equiv -z_c(x,H)$, trivially. The associated
LIM has point-disorder since $z_c(x,H)$ is delta-correlated in $x$ and $H$.
The LIM corresponding to the Manna model has a noise field $\eta(x,H)$
which is point-like and correlated: in the $H$-direction because
of random-walk like 
increments in $\eta(x,H)$  and in the $x$-direction
because of the short-ranged grain conservation
(if a nn of $x$ gained a grain when $H(x,t)$ increased other nn's did not)
\cite{prep}.

For non-SOC (periodic or open) boundary conditions the LIM
has a depinning transition at a critical force $F_c$. At the
transition, the scaling exponents of the LIM with columnar, point-like, 
and,  depending on the details, correlated disorder differ 
(see \cite{narayan-fisher:1993,nattermann-etal:1992} for
the effect of noise correlations). This means that 
avalanches have different spatial and temporal
properties since the critical exponents like the
{\it roughness exponent} $\chi$ of the LIM depend on the noise. 
In particular, the BTW model is in a different universality class from the
others \cite{avalanches,chessa-etal:1998} as it has no $\eta$-noise.
The LIM is invariant
to forces that are static in the $H$-direction \cite{narayan-fisher:1993} 
which makes $\eta$ a relevant perturbation. 
For the Manna-model, we conjecture that it may
be in the point-disorder LIM class despite the
correlations in $\eta$ \cite{narayan-fisher:1993,nattermann-etal:1992}.
This prediction seems to be shown to be true in $d=2$ 
in Ref.~\cite{new-av}.
The upper critical dimension of the LIM is $d_u = 4$ for all
these kinds of noise because of the Laplacian in Eq.~(\ref{eq:btw-equ}).

These conclusions are based on the continuum limit of the LIM.
However, sandpile models are discretized versions thereof, with
the additional $\sigma$-noise term in the equation. In the point-disorder 
LIM, various numerical approaches indicate that the $\sigma$ is irrelevant 
for the rice-pile model with periodic boundary conditions 
\cite{nattermann-etal:1992,leschhorn:1993}. This can
be understood by considering the sum of the $\eta$- and $\sigma$-noise
terms: the velocity constraint just renormalizes the $\eta$-field
since the velocity would in any case be of the order of unity,
or, the actual value of $\sigma$ depends on $z_c (x,H)$.

For the other scenarios one has to answer the question, whether
the $\sigma$-noise changes the universality class of the model at hand.
This would mean simply that the sandpile model can be a ``bad 
discretization''
of a continuum interface equation so that the $\sigma$-noise changes the
properties of the continuum model. Notice that the way $\sigma(x,H)$
becomes non-zero (in these models $\sigma \leq 0$) depends on the 
model. In the Manna-model a site with $z_x=1$ can also get two grains 
from the {\it same\/} nn leading to $\sigma \neq 0$.
This is point-like disorder since the decision to give two grains
has no correlations with any earlier events. Thus the Manna model
may be in the rice-pile universality class as noted above.
Like in the rice-pile model, the $\sigma$-noise should not have 
any strong correlations due to the randomness in the 
avalanches.

In the BTW and rice-pile models the $\sigma$-term 
arises (see again Fig.~\ref{fig:explain})) from the 
{\it simultaneous topplings\/} of nn's of site $j$. Thus
in $d>2$ the role of the noise is diminished, and in particular
it should not affect the upper critical dimension. 
Notice that the $\sigma$-noise changes off the critical
point (since more of the neighbors are likely to be active
at the same time), and may thus play a role
in systems off criticality like
in the 'fixed density' (or energy) ensemble corresponding to a constant 
total force \cite{dickman-etal:1998}. 
The BTW model has also the Abelian property \cite{dhar:1998} and
thus the $\sigma$-field of any particular sample
depends on the exact order of the topplings.

The $\sigma$-noise is next studied as such to elucidate
the role of the SOC boundary conditions in the interface
equation (open boundaries for the sandpile). We look
at (mostly) the BTW and the other models by numerical simulations 
in 2d in the normal SOC sandpile ensemble using parallel
dynamics for all active sites at each time step, as one
would do in an interface model. Quantities of
interest are the average of $\sigma$, and its spatial
dependence on $x$.
The $\sigma$-field is constructed from the relation
(\ref{f-renorm}).
%any of the avalanches taking a BTW model recurrent state to another.
%In the BTW model the $\sigma$-noise is 'self-generated'
%from the recurrent state by the selection of 
%the site on which a particle is deposited. 
%Since the model is 
%Abelian, two active nn sites can be toppled in arbitrary order

First we look at the finite size dependence of the probability
to have a non-zero $\sigma$, $P_{L}(\sigma < 0)$. It
can be studied as averaged over dissipating (grains leave
the system), and non-dissipating avalanches.
For all these $P_L$ increases with the system size $L$.
The finite size scaling  Ansatz $P_L \sim \left<P\right>_\infty + aL^{-b}$, 
fitted  to the data \cite{numbers}, gives 
correction exponents that are close to $b=2/3$ for
bulk and dissipating avalanches, and $-1$ for
all avalanches; $a$ is negative in all cases.
%Since the $\sigma$-noise is {\em not\/} constant a
%correction is induced to the 'driving force' 
%(since $\left<P\right> \neq 0$),
%and the fluctuations of $\sigma - \left<\sigma\right>$ are $L$-dependent. 
The asymptotic BTW values are $\left<P\right> \simeq 0.081$ for all
and $\left<P\right>_d \simeq 0.121$ for dissipating avalanches.
Thus the dissipating avalanches have typically
stronger $\sigma$-noise: the interface tries to move faster. 
They also contribute significantly into $\left<\sigma\right>$
since the total number of topplings is dominated by such avalanches. 
Therefore fluctuations of the $\sigma$-field
may be related to the fluctuations
in the average grain number,
$\left<z_x\right> \simeq \left<z_x\right>_c$. 
For the Manna and ricepile models
$\left<\sigma\right>$ also decreases with $L$ \cite{prep}.
%Fig.2

\begin{figure}%[htb]
\twofigures[width=7cm]{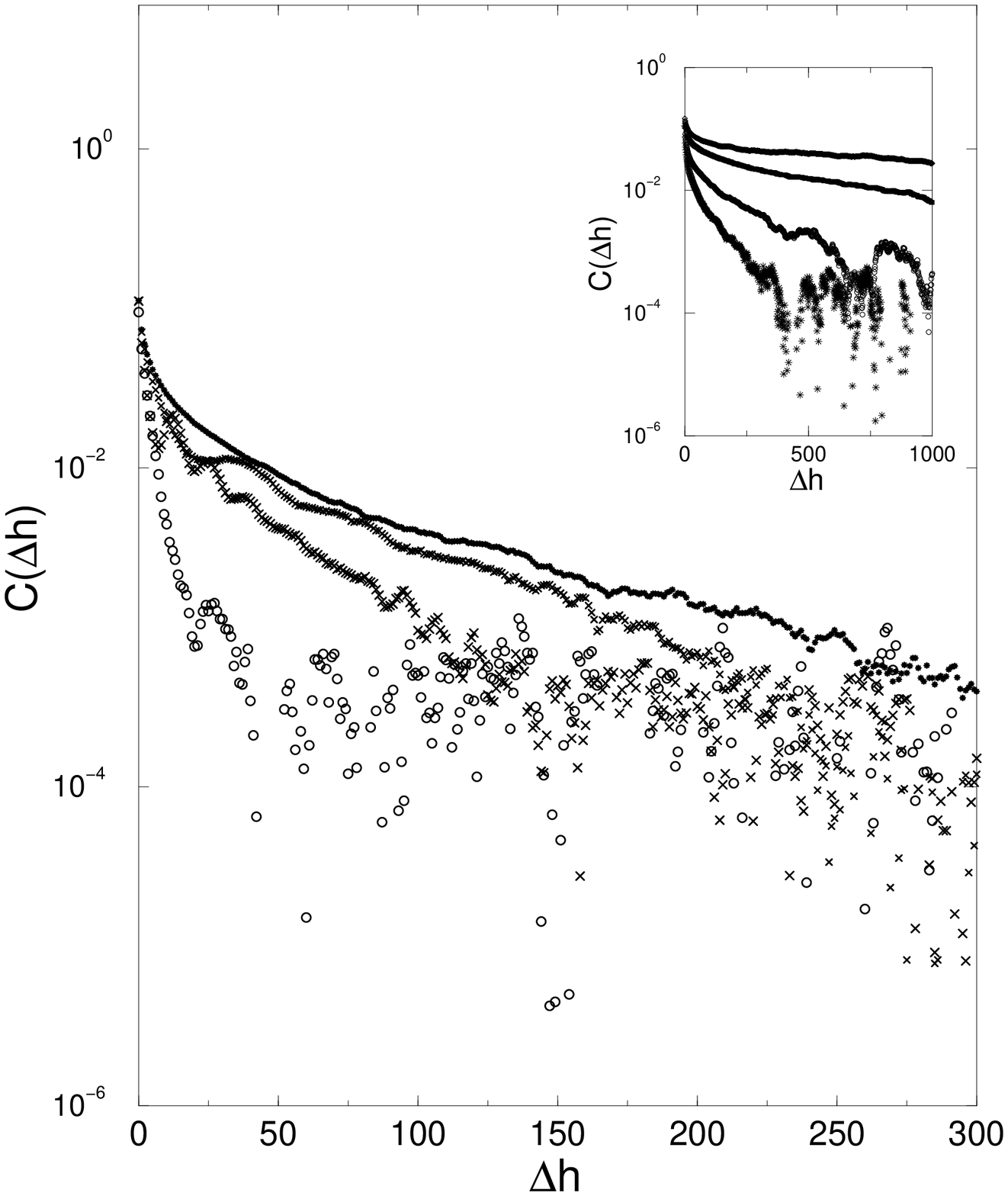}{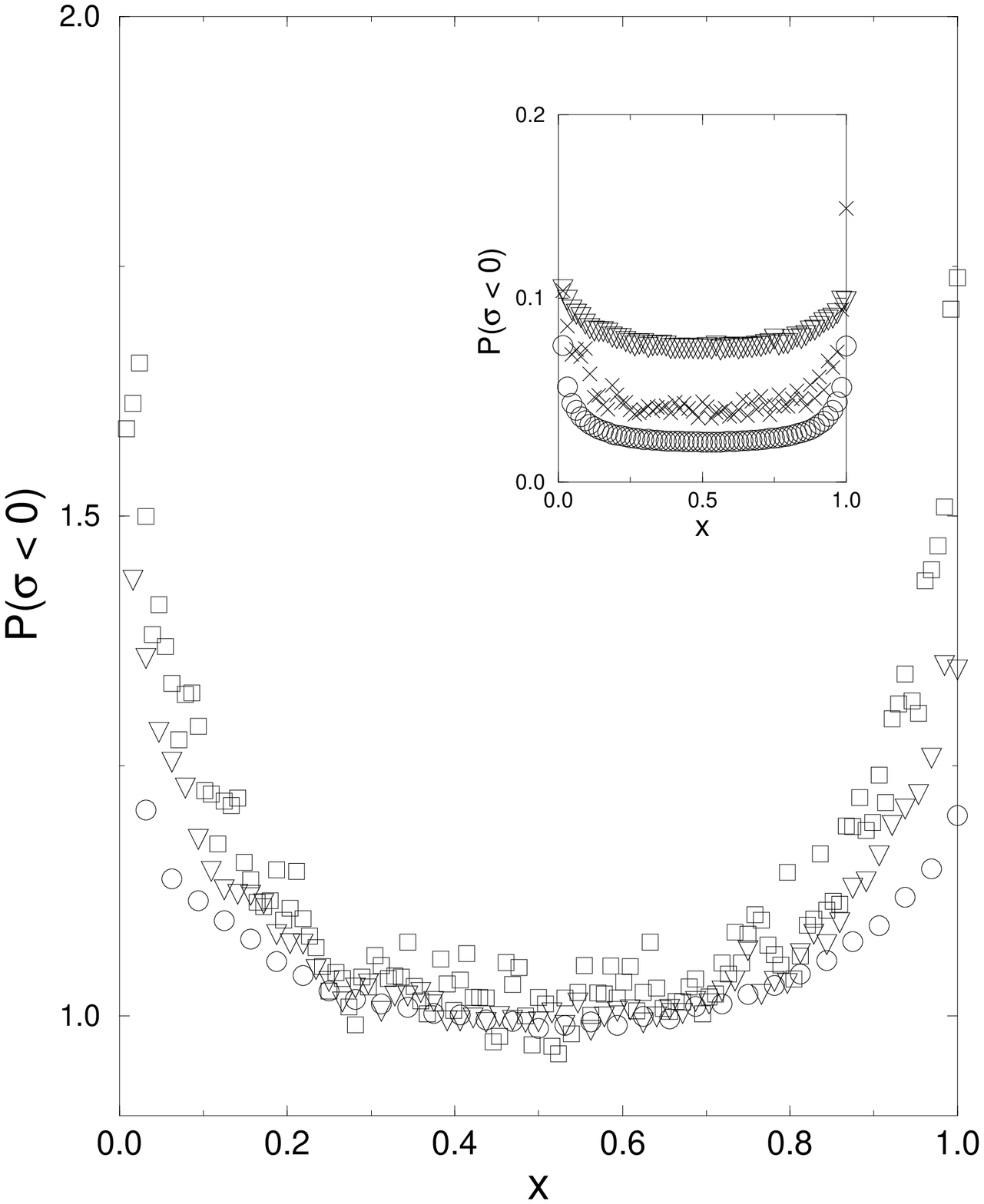}
\caption{Plots of $C(\Delta h)$ for BTW model noise at locations ranging
from the edge to the center (left to right) for $L=64$. 
Inset: $C(\Delta h)$ in
the center for $L=$ 64, 128, 256, 512 (bottom to top).}
\label{fig:corrf}
%\end{figure}

%\begin{figure}%[htb]
%\onefigure[width=7cm]{fig3new.ps}
\caption{$P(\sigma(x) <0)$ vs.\ $x$ scaled with $P(x=L/2)$ for various $L$ in
	the BTW model ($L=32$ (circles), $64$ (triangles), $128$
	(squares). Inset: for $L=64$ the same for BTW, Manna, 
        ricepile (triangles, circles, crosses).}
\label{fig:P-sigma}
\end{figure}

We checked the {\it spatial correlations\/} in  
$\sigma(x,H)$ by computing the two-point 
noise-noise correlation function 
$C(\delta H)=\left<\sigma(x,H+\delta H) \sigma(x,H) \right>
- \left< \bar{\sigma}^2\right>$, at a fixed $x = \rm{const.}$ \cite{xcorr}.
Figure~\ref{fig:corrf} demonstrates that
at the center and at the edges of the pile
the noise decays exponentially. The decay
is slower in the center the decay
lengths being in general proportional to $L$.
In the 'bulk' (see the second and third curve from left)
we see superposed on that behavior periodic oscillations.
Thus there is no spatial {\it translational invariance} 
\cite{drossel,ktitarev,barrat} since the correlations depend on $x$.
The details of the correlations of $\sigma$ in the BTW model may be 
related to observations of multiscaling in the BTW model \cite{Teb99} and 
manifest the columnar character of the $F$-noise.
In contrast, for the Manna/rice-pile models the correlations 
in $\sigma$ decay rapidly.

The average $\left< \sigma(x) \right>$ turns out to be non-uniform in $x$.
For instance, at the boundary the noise
is weaker since sites can receive grains only from $2d-1$ nn's.
The non-uniform $\sigma$ implies simply that the interface
tries to move faster in the center of the pile.
Next we investigate whether $\sigma(x,H) \neq 0$ 
if site $x$ does not topple any more during an avalanche.
In other words, is the velocity constraint related to the stopping
properties of avalanches? We look thus at the $\sigma$-noise
at the {\it elastic pinning paths\/} of an
interface model \cite{Makseetal}. The answer is relevant
for the existence of translational invariance as well.
Figure~\ref{fig:P-sigma} shows a data collapse of 
$P(\sigma(x,y) \neq 0)$
along the cut $y=L/2$, $1\leq x \leq L/2$,  scaled with $P(x\simeq L/2)$,
with the pinning-path constraint, i.e.\ the fraction
of all topplings with $\sigma <0$ such that the
toppling is the last at the site during an avalanche. 
The probability increases at the boundaries with 
system size $L$. This shows that the standard BTW sandpile has
no spatial translational invariance. This is true (see the inset) also 
for the Manna and rice-pile models. Their
pinning paths are determined by the 
configuration prior to the avalanche and the 
(point) disorder the avalanche encounters. In the BTW case the paths are
set by the original force/grain configuration as 
in a columnar-noise LIM, but not by the $\sigma$-noise field
itself as the Abelian character of the BTW also implies (the choice
of the toppling order does not change the final configuration).
It would be interesting to compare the BTW result to the wave picture 
of the BTW model \cite{dhar:1998,Pri96,Pac96}.
The lack of translational invariance is caused by the open 
boundary conditions that also imply a parabolic shape for
the interface and 
suggests that there is no
simple scaling as one would expect for a normal interface model,
in which case a relation $l^D \sim l^{d+\chi}$ would be valid
for the avalanche size vs.\ its linear dimension $l$. At most,
one should have $D_s \leq d+\chi$ for the cut-off dimension
of the probability distribution of the avalanche sizes, since
the open sandpile avalanches can of course not be 'over-critical'
with any effective $\chi' > \chi$. Note that this observation
seems to be true for any of the three models.
%since $\chi(x)$ depends on the model and $x$. 
%Indeed, for the BTW model at the boundaries $\chi=0$ due to the dissipation.
%Fig.3

In conclusion, sandpiles can be mapped to driven interfaces, 
by describing the dynamics with various types of quenched noise. 
The rice-pile model is equivalent to the random field linear interface model.
The BTW model has long-range on-site correlations that can be
studied via the $\sigma$-noise or the restriction $v \leq 1$.
It is because of the deposition noise $F(x,t)$ a columnar-noise 
LIM albeit with the velocity limitation. The Manna
model turns out to have ``correlated point-disorder''
and is thus in a different universality class from the BTW model.
The projection technique used for the Manna
toppling dynamics can be applied to e.g.\ the Zhang model, 
models with bulk dissipation of grains,
and the Olami-Feder-Christensen model 
\cite{Zhangetc,OFC,vespignani-zapperi:1997} and thus
further extensions of our work are certainly possible.
%For instance, in the last model without 
%conservation a finite correlation length appears \cite{prep}.
The discussion of the $\sigma$-term in the interface equation
for sandpiles makes it clear that studying the noise provides
a new tool for elucidating sandpile behavior and the role of
various boundary conditions in sandpiles.
The fundamental ingredients in sandpiles are slow drive and fast dissipation.
With open boundaries these combine to make the pile
spatially non-uniform, while the rules chosen are
reflected in the scaling of the self-organized critical state.

\acknowledgments
K. B. L. acknowledges the support by the Carlsberg foundation.

\newpage

\vfill
\vspace{4cm}


\begin{thebibliography}{99}
\bibitem{numbers} 
Typically 10$^7$ avalanches were collected
	for $L =$ 64, 128, 256, 512 for the FSS data.
\bibitem{xcorr} 
	The values of $\sigma$ with $H = \rm{const.}$ 
	become uncorrelated with increasing $H$ since the interface
	is not flat.
\end{thebibliography}
\end{document}